# Complex Dynamics of Real Nanosystems: Fundamental Paradigm for Nanoscience and Nanotechnology

A.P. Kirilyuk

*Institute of Metal Physics, 36 Vernadsky Av., Kiev-142, 03142 Ukraine*

Structure and dynamics of real nanosystems emerge from the unreduced solution of the underlying interaction problem. It has the property of dynamic multivaluedness giving genuine dynamic randomness and complexity. Chaoticity is irreducibly high in nanoscale systems and underlies exponentially high efficiency of their real, complex-dynamic operation. Applications include real, dynamically chaotic quantum devices, nano-biotechnology, and genetics.

Структура та динаміка реальних наносистем виникають як нередукований розв'язок проблеми взаємодії. Він має властивість динамічної багатозначеності, яка веде до справжньої динамічної випадковості та складності. В нанорозмірних системах хаотичність завжди є великою і пов'язана з експоненційно високою ефективністю їх реальної, складно-динамічної дії. Застосування включають реальні квантові прилади з хаотичною динамікою, нано-біотехнологію та генетику.

Структура и динамика реальных наносистем получены как нередуцированное решение проблемы взаимодействия. Оно обладает свойством динамической многозначности, дающим истинную динамическую случайность и сложность. В наноразмерных системах хаотичность всегда сильна и связана с экспоненциально высокой эффективностью их реальной, сложно-динамической работы. Приложения включают реальные квантовые устройства с хаотической динамикой, нано-биотехнологию и генетику.

Similar to any other kind of system, structure and operation of nanoscale systems is determined by underlying interaction processes. However, the role of detailed interaction development increases on those ultimately small scales, where system dimensions approach the *irreducible* size of *discrete* matter elements and correspondingly "coarse-grained" steps of their behaviour. Already general considerations show that this should lead to relative importance of discrete, intrinsically chaotic regimes of real interaction dynamics in nanosystem structure emergence and operation [1], as opposed to various generally "smooth", mean-field, unitary evolution models usually applied to larger-scale systems. In this report we present a first-principle analysis of real nanosystem behaviour, based on the unreduced, nonperturbative solution of the underlying (arbitrary) many-body interaction problem that leads to the universal concept of dynamic complexity/chaoticity [1-7] and genuine, fundamentally substantiated nanoscience paradigm [1].

We start with a unified formulation of arbitrary (generic) interaction problem in terms of *existence equation* representing a generalised version of particular dynamic equations (e.g. Schrödinger equation) and actually expressing only the fact of unreduced interaction as such:

$$\left\{ \sum_{k=0}^{N} \left[ h_k(q_k) + \sum_{l>k}^{N} V_{kl}(q_k, q_l) \right] \right\} \Psi(Q) = E\Psi(Q), \qquad (1)$$

where $h_k(q_k)$ is the generalised Hamiltonian of the $k$-th component with the degrees of freedom $q_k$, $V_{kl}(q_k, q_l)$ is the interaction potential between the $k$-th and $l$-th components, $\Psi(Q)$ is the system state-function, $Q \equiv \{q_0, q_1, ..., q_N\}$, $E$ is the generalised Hamiltonian eigenvalue, and summations are performed over all ($N$) system components. The Hamiltonian equation form does not play the key role as such and is chosen because it can be rigorously derived as indeed universal expression of system dynamics [1,2,4,7], corresponding to an observable measure of dynamic complexity defined below. We can rewrite eq. (1) in a more convenient form, reflecting the fact that one of the degrees of freedom, for example $q_0 \equiv \xi$, is physically separated from other ones, since it serves as a common, distributed system measure or interaction entity, such as space coordinate of system elements or input/output interaction field:

$$\left\{ h_0(\xi) + \sum_{k=1}^{N}[h_k(q_k) + V_{0k}(\xi, q_k)] + \sum_{l>k}^{N} V_{kl}(q_k, q_l) \right\} \Psi(\xi, Q) = E\Psi(\xi, Q), \quad (2)$$

where $Q \equiv \{q_1,...,q_N\}$ and $k, l \geq 1$ here and below.

We proceed with problem expression in terms of eigenfunctions $\{\varphi_{kn_k}(q_k)\}$ and eigenvalues $\{\varepsilon_{n_k}\}$ of non-interacting components:

$$h_k(q_k)\varphi_{kn_k}(q_k) = \varepsilon_{n_k}\varphi_{kn_k}(q_k), \quad (3)$$

$$\Psi(\xi, Q) = \sum_{n \equiv (n_1, n_2, ..., n_N)} \psi_n(\xi) \Phi_n(Q), \quad (4)$$

where $\Phi_n(Q) \equiv \varphi_{1n_1}(q_1)\varphi_{2n_2}(q_2)...\varphi_{Nn_N}(q_N)$ and $n \equiv (n_1, n_2, ..., n_N)$ takes all possible eigenstate combinations. Inserting eq. (4) into eq. (2) and performing eigenfunction separation (e.g. by a scalar product operation), we obtain the system of equations for $\psi_n(\xi)$ equivalent to eqs. (1), (2):

$$[h_0(\xi) + V_{nn}(\xi)]\psi_n(\xi) + \sum_{n' \neq n} V_{nn'}(\xi)\psi_{n'}(\xi) = \eta_n \psi_n(\xi), \quad (5)$$

where

$$\eta_n \equiv E - \varepsilon_n, \quad \varepsilon_n \equiv \sum_k \varepsilon_{n_k}, \quad V_{nn'}(\xi) = \sum_k \left[ V_{k0}^{nn'}(\xi) + \sum_{l>k} V_{kl}^{nn'} \right], \quad (6)$$

$$V_{k0}^{nn'}(\xi) = \int_{\Omega_Q} dQ \Phi_n^*(Q) V_{k0}(q_k, \xi) \Phi_{n'}(Q), \quad V_{kl}^{nn'}(\xi) = \int_{\Omega_Q} dQ \Phi_n^*(Q) V_{kl}(q_k, q_l) \Phi_{n'}(Q). \quad (7)$$

If we try to "solve" the nonintegrable eqs. (5) by substitution of variables using Green function technique, we obtain a dynamically rich problem formulation in terms of *effective potential (EP) formalism* and *qualitatively new* solution to its *unreduced* version that contains *many* locally complete, and therefore *incompatible*, versions of system configuration, or *realisations*, forced by the *same* interaction to *permanently* replace each other in a *dynamically chaotic* order (*dynamic multivaluedness, or redundance*, phenomenon) [1-7]:

$$\rho(\xi, Q) \equiv |\Psi(\xi, Q)|^2 = \sum_{r=1}^{N_\Re} {}^\oplus \rho_r(\xi, Q), \quad (8)$$

where $\rho(\xi, Q)$ is the measured density, $\rho_r(\xi, Q) \equiv |\Psi_r(\xi, Q)|^2$ is the *r*-th realisation density, and the *dynamically probabilistic* sum, marked by $\oplus$, describes the unceasing chaotic change of all ($N_\Re$) system realisations. The state-function $\Psi_r(\xi, Q)$ for the *r*-th realisation is obtained in the form:

$$\Psi_r(\xi, Q) = \sum_i c_i^r \left[ \Phi_0(Q)\psi_{0i}^r(\xi) + \sum_{n,i'} \frac{\Phi_n(Q)\psi_{ni'}^0(\xi) \int_{\Omega_\xi} d\xi' \psi_{ni'}^{0*}(\xi') V_{n0}(\xi')\psi_{0i}^r(\xi')}{\eta_i^r - \eta_{ni'}^0 - \varepsilon_{n0}} \right], \quad (9)$$

where $n \neq 0$, $\varepsilon_{n0} \equiv \varepsilon_n - \varepsilon_0$, $c_i^r$ are coefficients giving causal Born's rule for realisation probabilities [1,2], and $\{\psi_{0i}^r(\xi), \eta_i^r\}$ are the *r*-th realisation eigen-solutions of the *effective* existence equation:

$$h_0(\xi)\psi_0(\xi) + V_{\text{eff}}(\xi; \eta)\psi_0(\xi) = \eta\psi_0(\xi), \quad (10)$$

the EP operator $V_{\text{eff}}(\xi; \eta)$ being defined by its action:

$$V_{\text{eff}}(\xi;\eta_i^r)\psi_{0i}^r(\xi) = V_{00}(\xi)\psi_{0i}^r(\xi) + \sum_{n,i'} \frac{V_{0n}(\xi)\psi_{ni'}^0(\xi) \int_{\Omega_\xi} d\xi' \psi_{ni'}^{0*}(\xi')V_{n0}(\xi')\psi_{0i}^r(\xi')}{\eta_i^r - \eta_{ni'}^0 - \varepsilon_{n0}} \quad , \quad (11)$$

and $\{\psi_{ni}^0(\xi), \eta_{ni}^0\}$ are eigen-solutions of a truncated system (where $n, n' \neq 0$, contrary to eqs. (5)):

$$[h_0(\xi) + V_{nn}(\xi)]\psi_n(\xi) + \sum_{n' \neq n} V_{nn'}(\xi)\psi_{n'}(\xi) = \eta_n \psi_n(\xi). \quad (12)$$

While dynamic redundance and chaotic realisation change are due to the *essentially nonlinear* dependence of the unreduced EP, eqs. (10)-(11), on the solutions to be found, application of the same EP method to eqs. (12) reveals the *probabilistically fractal* structure of system dynamics [1,2,5,7], forming the truly *complete general solution* of eq. (8). The latter is accompanied by the values of *a priori realisation probabilities*, $\alpha_r$:

$$\alpha_r = \frac{N_r}{N_\Re} \left( N_r = 1, ..., N_\Re; \sum_r N_r = N_\Re \right), \quad \sum_r \alpha_r = 1 \quad , \quad (13)$$

where $N_r$ is the number of "elementary" realisations forming the $r$-th compound realisation [1-5]. Realisation probabilities obey also the *causal Born rule*, $\alpha_r = |\Psi(X_r)|^2$, where $X_r$ is the $r$-th realisation configuration and $\Psi(x)$ the *generalised wavefunction*, or *intermediate* ("main") system realisation, satisfying the *generalised Schrödinger equation* [1,2,4]. Note that usual, perturbative models imply *rejection* of *all* system realisations but a *single*, "averaged" one. We call the ensuing reduction *dynamically single-valued, or unitary, solution* and approach (of the canonical theory). It corresponds to *zero* value of the *unreduced dynamic complexity*, $C$, since the latter is *universally* defined [1-7] as any growing function of the total number of system realisations, or rate of their change, equal to zero for the case of only one realisation: $C = C(N_\Re)$, $dC/dN_\Re > 0$, $C(1) = 0$.

Due to the intrinsic universality of the unreduced EP analysis, we can classify *all* possible regimes of real, complex interaction dynamics into two main groups, or limiting cases, the *uniform, or global, chaos* and *multivalued self-organisation*, passing to one another with parameter change and alternating in real interaction development [1-3]. Uniform chaos emerges for close enough parameters of system components (universally represented by their main frequencies) and is characterised by close probabilities of sufficiently different realisations. The opposite case of multivalued self-organisation is due to essentially different component properties, where slowly moving, "inertial" component(s) tend to confine a multitude of quickly and chaotically changing realisations. It is this latter limit that gives rise to the unitary model reduction, including the usual "self-organisation". The real, dynamically multivalued structure *always* contains genuine dynamical chaos [1-3], although it can remain unnoticed behind the quasi-regular external shape. The unreduced self-organisation regime unifies essentially extended versions of various separated models of the unitary "science of complexity", such as "self-organisation", "self-organised criticality", "control of chaos", "synchronisation", and "phase locking".

The unreduced EP formalism leads to the *universal criterion of transition* between the multivalued self-organisation (general "regularity") and global chaos, occurring at equal characteristic frequencies, or energy eigenvalue separations, of system components [1-3]. The strong, uniform chaos observed around that *resonance condition* gradually passes to ever more regular structures outside of resonance. This universal criterion of chaos leads to the major conclusion about *impossibility* of any regular enough dynamics in *ultimately small*, nanoscale systems [1]. Indeed, their component eigenvalue separations and frequencies tend to be comparable simply because they approach the unique limit of just one bound (excited) state. The same condition can be expressed in terms of strongly discrete component sizes (related to frequencies): in a few-atom system the minimum discrete change is always relatively big and *necessarily random*, as we can see now from the *unreduced* interaction analysis. In other words, a real nanosystem is always close to the resonant, global-chaos dynamics, since there is only *strictly limited* "room at the bottom", in accord with the undeniable quantum postulates [1] and contrary to the popular, but provably *wrong* slogan of the unitary theory [8] presented as nanoscience foundation.

Moreover, the intrinsic chaoticity of nanosystem dynamics is its essential advantage, rather than deficiency, since useful operation power of unreduced, complex-dynamical interaction process is *exponentially higher* than that of its unitary, regular, sequential model [1]. It follows from the fact that the efficiency of *self-multiplying* fractal structure of unreduced interaction is given by the number of all combinations of links between system elements, or $N! \simeq \sqrt{2\pi N}(N/e)^N \sim N^N$ for $N$ links. Since $N$ is usually a large number, one obtains really huge, practically infinite values of complex-dynamic system efficiency with respect to any unitary model whose efficiency grows only as $N^\beta$ ($\beta \sim 1$). In fact, the relative system power is given by its unreduced dynamic complexity proportional to the link combination number, which explains the force of nonunitary operation. The unavoidable payment for this advantage is chaoticity, making impossible any usual, regular "programming" and control, but providing a possibility of *independent, quasi-autonomous* system operation and *purposeful self-development*, lacking in any unitary operation scheme.

This combination of chaoticity, efficiency, and creativity of real nanosystem dynamics forms the genuine, fundamentally substantiated *nanoscience and nanotechnology paradigm* as that of *essentially complex-dynamic (multivalued, chaotic) structure emergence and operation on ultimately small scales*. This rigorously derived paradigm and *definition* of nanoscience is *already realised* in all *natural* "nanodevices" in living organisms, explaining their "miraculous" properties. Particular applications include [1,2,5,7]: (1) *real, dynamically chaotic quantum devices* that *cannot* function at all in their unitary version; (2) "difficult" cases of irreducibly *strong interaction* in solid-state physics, such as quantum (wave) behaviour of many-atom systems and high-temperature superconductivity; (3) full-scale, sustainable nanomachines and realistic *nano-biotechnology* using partially natural or totally artificial structures; (4) reliable, *provably constructive genetics*; and (5) understanding and development of natural and *true* artificial *intelligence* and *consciousness*.

# Supplement:

# Brief (poster) presentation of the report at NANSYS 2004 (Kyiv, 12-14 October 2004)

# Complex Dynamics of Real Nanosystems: Fundamental Paradigm For Nanoscience And Nanotechnology

Andrei P. Kirilyuk

Institute of Metal Physics, 03142 Ukraine

kiril@metfiz.freenet.kiev.ua

http://myprofile.cos.com/mammoth

Structure and dynamics of real nanosystems emerge from the unreduced solution of the underlying interaction (many-body) problem. It has the property of dynamic multivaluedness giving genuine dynamic randomness and complexity. Chaoticity is irreducibly high in nanoscale systems and underlies exponentially high efficiency of their real, complex-dynamic operation. Applications include real quantum devices, nano-biotechnology, and genetics.

Структура и динамика реальных наносистем получены как нередуцированное решение проблемы взаимодействия (многих тел). Оно обладает свойством динамической многозначности, дающим истинную динамическую случайность и сложность. В наноразмерных системах хаотичность всегда сильна и связана с экспоненциально высокой эффективностью их реальной, сложно-динамической работы. Приложения включают реальные квантовые устройства, нано-биотехнологию и генетику.

# Сложная динамика реальных наносистем:
## Фундаментальная парадигма для нано-науки и технологии

А.П. Кирилюк

Институт металлофизики НАН Украины, Киев

Динамика реальных наносистем рассматривается в рамках непертурбативного анализа проблемы многих тел произвольной конфигурации. Получено полное решение уравнения общего вида, описывающего такую систему, которое избегает обычных приближений теории возмущений или "точно решаемых" моделей. В результате обнаружен эффект динамической многозначности: любое реальное взаимодействие даёт много одинаково возможных, но взаимно несовместимых конфигураций, или "реализаций", системы, которые вынуждены постоянно сменять друг друга в динамически случайном порядке, обуславливая внутреннюю хаотичность и универсально определённую сложность возникающей структуры и поведения системы. Системы предельно малых размеров соответствуют нижайшим, квантовым и первым классическим, уровням сложности мира, где доминирует один из двух предельных режимов сложной динамики, режим однородного, или глобального, хаоса. Он возникает в случае резонанса между основными частотами системы (неизбежного при этих предельно малых размерах) и характеризуется относительно частыми переходами между реализациями системы, которые заметно отличаются друг от друга (т. е. имеет место сильно выраженная нерегулярность). Эта неизбежность сложного, динамически многозначного поведения наноразмерных систем, полученная на основе нередуцированного решения соответствующих уравнений общего вида, предлагается в качестве фундаментальной основы для нано-науки и технологии, которая раскрывает существенное, качественное отличие поведения наносистем от более крупных структур (в отличие от динамически однозначных, внутренне регулярных моделей и подходов к нанотехнологии, которые необоснованно переносят одни и те же, сильно редуцированные решения на новый пространственный масштаб). Эта фундаментальная, строго обоснованная парадигма нанотехнологии показывает, что с уменьшением масштаба неизбежен переход к качественно новому типу устройств, которые напоминают биологические наносистемы, с их хаотической динамикой, постоянным возникновением новых структур и переходами между квантовыми и классическими уровнями (сложной) динамики. Существенными преимуществами такого типа устройств, выведенными в рамках предлагаемой концепции сложности, являются их экспоненциально высокая эффективность (по сравнению с обычной техникой регулярного действия) и внутренний творческий потенциал. Мы описываем различные приложения такой каузально полной концепции сложно-динамической нанотехнологии, включая (истинный) квантовый хаос, динамически объяснённый переход от квантового к классическому поведению, проблему (реальных) квантовых компьютеров, сложно-динамическую основу многочастичных квантовых состояний и других "трудных" случаев конденсированного состояния, проблемы генетики, нано-биотехнологии и искусственного интеллекта и сознания.

Generalised dynamic equation for many-body interaction problem:

$$\left\{\sum_{k=0}^{N}\left[h_k(q_k)+\sum_{l>k}^{N}V_{kl}(q_k,q_l)\right]\right\}\Psi(Q)=E\Psi(Q)$$

or

$$\left\{h_0(\xi)+\sum_{k=1}^{N}\left[h_k(q_k)+V_{0k}(\xi,q_k)\right]+\sum_{l>k}^{N}V_{kl}(q_k,q_l)\right\}\Psi(\xi,Q)=E\Psi(\xi,Q)$$

The unreduced (nonperturbative) general solution is *always probabilistic* (phenomenon of *dynamic multivaluedness = intrinsic chaoticity*):

$$\rho(\xi,Q)=\sum_{r=1}^{N_\Re}{}^\oplus\rho_r(\xi,Q)$$

⊕ designates *dynamically probabilistic* sum over *incompatible* realisations

Two limiting regimes of complex dynamics:
*multivalued self-organisation* (or SOC) and *uniform (global) chaos*

Universal criterion of global (strong) chaos:

$$\kappa\equiv\frac{\Delta\eta_i}{\Delta\eta_n}=\frac{\omega_\xi}{\omega_Q}\simeq 1$$

or *resonance* between the main system motions

Criterion of quasi-regularity (self-organisation): $\kappa\ll 1$ or $\kappa\gg 1$.

In *genuine nanosystems* (ultimately small sizes) *always* $\kappa\sim 1$
and therefore **one can only have strong (and true) chaos** regime:

Coarse-grained interaction dynamics cannot be close to regularity

Ordinary, unitary dynamic "models" are inapplicable

let's transform the unitary model *defect* (system failure)
into the complex-dynamic system *advantage* (superior power and qualities)

**Essentially chaotic dynamics gives *huge advantage in efficiency***

***Chaotic*** network efficiency is determined
by the number of ***all combinations of links***, or

$$N! \simeq \sqrt{2\pi N}(N/e)^N \sim N^N$$

for *N* links, where *N* is usually large itself in a full-scale system

One obtains <u>exponentially big advantage with respect to unitary efficiency</u>
growing only as $N^\beta$ ($\beta \sim 1$)

This huge advantage is the expression of ***autonomous creativity***
of unreduced (multivalued) interaction dynamics,
with the <u>inevitable</u> "payment" for it by <u>chaoticity</u> of results

---

Intrinsically (and strongly) <u>chaotic</u> operation
with <u>huge efficiency</u> and <u>autonomous creativity</u>
is the new, genuine, <u>fundamentally substantiated</u>
***paradigm of nanoscience and nano-biotechnology***

---

There is ***no*** "plenty of room at the bottom" (cf. R. Feynman [8]):
<u>the unitary nanoscience hype is fundamentally wrong and misleading</u>
(тобто брехня у кубе!)

<u>Applications of genuine nanoscience paradigm</u> (already being realised):

(1) *real quantum devices* (they *cannot* function in their unitary version)
→ <u>genuine quantum chaos</u>

(2) "difficult" cases of *strong interaction* in solid-state physics
→ quantum behaviour of many-atom systems and HTSC

(3) full-scale, sustainable nanomachines and realistic *nano-biotechnology*

(4) reliable, *provably constructive genetics*

(5) understanding and development of natural
and *true* artificial *intelligence* and *consciousness*

# APPENDIX: Mathematical Details

## From Perturbative Model Reduction to Unreduced Problem Solution

Arbitrary interaction process (system) dynamics in terms of the (free) component eigen-modes:

$$h_0(\xi)\psi_n(\xi) + \sum_{n'} V_{nn'}(\xi)\psi_{n'}(\xi) = \eta_n \psi_n(\xi) \ , \tag{a}$$

where the total system state-function is

$$\Psi(q_0, q_1, ..., q_N) \equiv \Psi(\xi, Q) = \sum_{n \equiv (n_1, n_2, ..., n_N)} \psi_n(q_0)\varphi_{1n_1}(q_1)\varphi_{2n_2}(q_2)...\varphi_{Nn_N}(q_N) \equiv \sum_n \psi_n(\xi)\Phi_n(Q) . \tag{b}$$

After finding eigen-solutions of the system of equations (A), $\{\psi_{ni}(\xi), \eta_{ni}\}$, the *general solution* of a problem is their combination, such as

$$\Psi(\xi, Q) = \sum_n c_i \psi_{ni}(\xi)\Phi_n(Q) . \tag{c}$$

In reality $\{\psi_{ni}(\xi), \eta_{ni}\}$ are always found from a perturbative approximation:

$$\left[ h_0(\xi) + V_{nn}(\xi) + \tilde{V}_n(\xi) \right]\psi_n(\xi) = \eta_n \psi_n(\xi) \ , \quad V_0(\xi) < \tilde{V}_n(\xi) < \sum_{n'} V_{nn'}(\xi) . \tag{d}$$

The unreduced general solution of the same problem
is the *dynamically probabilistic* sum over *redundant* system realisations:

$$\rho(\xi, Q) \equiv |\Psi(\xi, Q)|^2 = \sum_{r=1}^{N_\Re} {}^\oplus \rho_r(\xi, Q) \ , \quad \rho_r(\xi, Q) = |\Psi_r(\xi, Q)|^2 \ , \tag{A.1}$$

$$\Psi_r(\xi, Q) = \sum_i c_i^r \left[ \Phi_0(Q)\psi_{0i}^r(\xi) + \sum_{n,i'} \frac{\Phi_n(Q)\psi_{ni'}^0(\xi)\int_{\Omega_\xi} d\xi' \psi_{ni'}^{0*}(\xi')V_{n0}(\xi')\psi_{0i}^r(\xi')}{\eta_i^r - \eta_{ni'}^0 - \varepsilon_{n0}} \right] , \tag{A.2}$$

where $\{\psi_{0i}^r(\xi), \eta_i^r\}$ are eigen-solutions of the *effective* equation:

$$h_0(\xi)\psi_0(\xi) + V_{\text{eff}}(\xi; \eta)\psi_0(\xi) = \eta \psi_0(\xi) \ , \tag{B.1}$$

$$V_{\text{eff}}(\xi; \eta_i^r)\psi_{0i}^r(\xi) = V_{00}(\xi)\psi_{0i}^r(\xi) + \sum_{n,i'} \frac{V_{0n}(\xi)\psi_{ni'}^0(\xi)\int_{\Omega_\xi} d\xi' \psi_{ni'}^{0*}(\xi')V_{n0}(\xi')\psi_{0i}^r(\xi')}{\eta_i^r - \eta_{ni'}^0 - \varepsilon_{n0}} \ , \tag{B.2}$$

and $\{\psi_{ni}^0(\xi), \eta_{ni}^0\}$ are eigen-solutions of the truncated system of equations:

$$h_0(\xi)\psi_n(\xi) + \sum_{n'} V_{nn'}(\xi)\psi_{n'}(\xi) = \eta_n \psi_n(\xi) \ , \quad \underline{n, n' \neq 0} \ . \tag{C}$$

Elementary length $\Delta x = \lambda = \Delta \eta_i^r$, time $\Delta t = \Delta x / v_0$, action $A_0 = V_{\text{eff}} \Delta t$ \qquad (D)

# Dynamic Redundance (Multivaluedness) as the Origin of Causal Randomness in Arbitrary System with Interaction. Dynamic Entanglement of Interacting Entities & Fractality

Interaction between two many-body ('many-point') entities (objects):

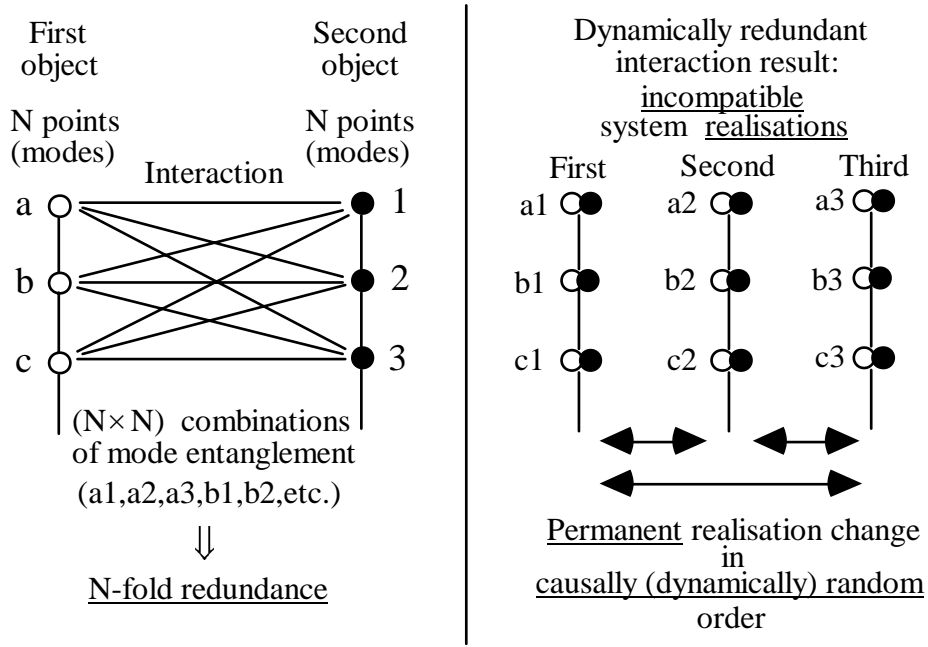

Permanent dynamic instability in any system with interaction by dynamic feedback loops in the unreduced interaction development:

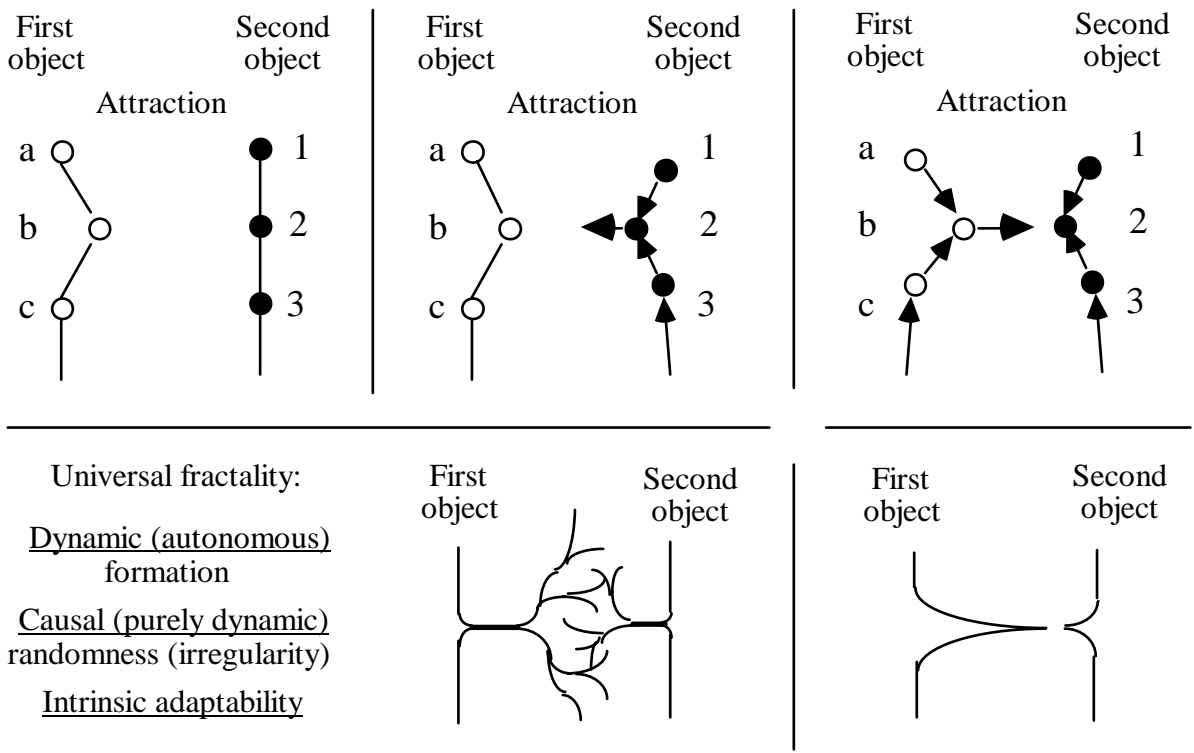